\documentstyle[12pt,a4,epsf]{article}

\newcommand{\wt}{\widetilde}
\newcommand{\ol}{\overline}

\begin{document}

\begin{titlepage}
\title{
\hfill\parbox{4cm}
{\normalsize UT-920\\{\tt hep-th/0012254}\\December, 2000}\\
\vspace{2cm}
A D2-brane realization of Maxwell-Chern-Simons-Higgs systems
\vspace{1cm}
}
\author{
Yosuke Imamura\thanks{{\tt imamura@hep-th.phys.s.u-tokyo.ac.jp}}
\\[7pt]
{\it Department of Physics, University of Tokyo, Tokyo 113-0033, Japan}
\vspace{2cm}
}
\date{}

\maketitle
\thispagestyle{empty}

\begin{abstract}
\normalsize
We show that ${\cal N}=2$ supersymmetric
Maxwell-Chern-Simons-Higgs systems in three dimension
can be realized as gauge theories on a D2-brane in D8-branes background
with a non-zero $B$-field.
We reproduce a potential of Coulomb branch of the Chern-Simons theory
as a potential of a D2-brane in a classical D8-brane solution
and show that each Coulomb vacuum is realized by
a D2-brane stabilized in the bulk at a certain distance from
D8-branes.
\end{abstract}

\end{titlepage}
\section{Introduction}
Chern-Simons theories in three-dimension\cite{dunne} have
many interesting properties:
rich phase structures, existence of solitons, etc.
In this paper, we discuss their brane realization.

One way to realize three-dimensional gauge theories in string theory is
to use Hanany-Witten type configurations.
${\cal N}=4$ gauge theories are realized on
D3-branes stretched between parallel NS5-branes\cite{HW}.
We can decrease the number of supersymmetries
and can introduce a non-zero Chern-Simons coupling
by replacing NS5-branes by $(p,q)$ 5-branes
and putting them at appropriate angles\cite{ohta1,ohta2,ohta3,ohta4,BHKK,ohta5}.
Using these configurations,
moduli space, phase structures and solitons of the theories
are studied in geometrical ways.

We can also realize three-dimensional gauge theories
as theories on D2-branes\cite{SeibergThreeDim,SW96}.
For example, an ${\cal N}=4$ $U(1)$ gauge theory with $N_f$ flavors
is realized on a D2-brane in $N_f$ D6-branes background.
However, unlike the Hanany-Witten type configurations,
we come up against a problem
if we attempt to introduce a Chern-Simons term.
The D2-brane action involves the following term\cite{massived2}
\begin{equation}
S=\frac{\Lambda}{4\pi}\int A\wedge dA,
\end{equation}
where $\Lambda$ is a cosmological constant quantized in integer units.
Because we need to use this coupling
to introduce a nonzero Chern-Simons coupling,
we should put D2-branes in a massive IIA or, equivalently, a D8-brane background.
If we put a D2-brane parallel to a D8-brane, however,
the configuration becomes non-BPS and unstable.
In order to keep supersymmetries unbroken,
we have to take one direction on the D2-brane to be a dilatonic direction
perpendicular to the D8-brane.
Although ${\cal N}=(8,0)$ supersymmetry is realized in this case,
the three-dimensional Poincar\'e invariance is broken\cite{KR}.

In \cite{itoyama} open string spectra and BPS conditions of D$p$-D$p'$
systems are studied and it is pointed out that
D$p$-D($p+6$) systems can be made BPS by
turning on a constant $B$-field.
Recently, Mihailescu, Park and Tran\cite{MPT} and Witten\cite{d0d6}
take advantage of this fact to study gauge theories on D-branes.
In this paper, inspired by their idea,
we study a relation between
a D2-D8 configuration with a non-zero $B$-field
and a Maxwell-Chern-Simons-Higgs system
expected to be realized on the D2-brane.

\section{Maxwell-Chern-Simons-Higgs systems}
We consider a D2-D8 system with a non-zero $B$-field.
Let us assume that $N_f$ D8-branes are located at $x^9=q_I$
($I=1,2,\ldots,N_f$)
and one D2-brane spreads along the $x^0$, $x^1$ and $x^2$ directions.
We turn on the following components of the $B$-field.
\begin{equation}
\frac{B_{36}}{T}=\tan\theta_1,\quad
\frac{B_{47}}{T}=\tan\theta_2,\quad
\frac{B_{58}}{T}=\tan\theta_3,\quad
(0\leq\theta_i\leq\pi),
\label{bistan}
\end{equation}
where $T=1/(2\pi l_s^2)$ is the string tension.
This configuration is T-dual to
a D0-D6 system in \cite{itoyama,d0d6}
and a D3-D9 system in \cite{MPT}.
Therefore, we can divert results there
to analysis of our D2-D8 system.
A BPS condition for this system is
\begin{equation}
e^{i(\theta_1+\theta_2+\theta_3)}=-i.
\label{BPS}
\end{equation}
We define a parameter $r$ by
\begin{equation}
r=\theta_1+\theta_2+\theta_3-\frac{3\pi}{2}.
\end{equation}
This parameter corresponds to the $r$ in \cite{d0d6}.
We have fixed some ambiguities for definitions of
parameters differently from \cite{d0d6}
for later convenience.
Near the BPS point $r=0$, an ${\cal N}=2$ $U(1)$
gauge theory is
expected to be realized on the D2-brane.
The field content of this theory is
one $U(1)$ vector multiplet $(a,A_\mu,\lambda,D)$ from 2-2 strings
and $N_f$ chiral multiplets $(\phi_I,\psi_I,F_I)$ ($I=1,2,\ldots,N_f$)
from 2-8 strings.
Precisely, there are three more chiral multiplets from 2-2 strings
representing fluctuations of the D2-brane parallel to the D8-branes.
Because they are neutral under the $U(1)$ and decoupled
from other fields,
we shall neglect them in what follows.

Let us consider an ${\cal N}=2$ supersymmetric action
for these fields.
Kinetic terms of vector and chiral multiplets are
\begin{eqnarray}
{\cal L}_{\rm vector-kin}
&=&
\frac{1}{g^2}\left(-\frac{1}{4}F_{\mu\nu}F^{\mu\nu}
-\frac{1}{2}\partial_\mu a\partial^\mu a
+\frac{1}{2}D^2\right),
\label{d3n2gauge}\\
{\cal L}_{\rm chiral-kin}
&=&\frac{1}{g^2}\sum_{I=1}^{N_f}\left(-{\cal D}_\mu\ol\phi_I{\cal D}^\mu\phi_I
 +\ol F_IF_I+D|\phi_I|^2+a\ol F_I\phi_I+a\ol\phi_IF_I\right).
\end{eqnarray}
(We are now interested in only bosonic fields.)
For chiral multiplets, we can add mass terms
\begin{equation}
{\cal L}_{\rm chiral-mass}
=-\frac{1}{g^2}\sum_{I=1}^{N_f}M_I(\phi_I\ol F_I+\ol\phi_IF_I).
\end{equation}
Because gauge group is $U(1)$, we can introduce a Fayet-Iliopoulous term
\begin{equation}
{\cal L}_{\rm FI}=\frac{\xi}{g^2}D.
\end{equation}
Furthermore, in three-dimension, a Chern-Simons term is allowed.
\begin{equation}
{\cal L}_{\rm CS}
=\frac{k}{4\pi}\epsilon^{\alpha\beta\gamma}A_\alpha\partial_\beta A_\gamma,
\end{equation}
where $k$ is an integer-valued quantized Chern-Simons coupling.
The equation of motion for the gauge field is
\begin{equation}
\partial_\mu F^{\mu\lambda}
   +\frac{\mu}{2}\epsilon^{\lambda\mu\nu}F_{\mu\nu}=\mbox{current},\quad
\left(\mu=\frac{kg^2}{2\pi}\right).
\end{equation}
Because the second term on the left hand side plays a role
of a `mass term' of the gauge field,
we should introduce mass terms also for other component fields
in the vector multiplet
in order to make the action supersymmetric.
\cite{LLW,Ivanov}
\begin{equation}
{\cal L}_{\rm vector-mass}
=-\frac{\mu}{g^2}aD
\end{equation}

Gathering all and
eliminating auxiliary fields $F_I$ and $D$,
we obtain
\begin{equation}
V(a,\phi_I)=\frac{1}{g^2}\left[\frac{1}{2}(\sum_{I=1}^{N_f}|\phi_I|^2-\mu a+\xi)^2
                                    +\sum_{I=1}^{N_f}(a-M_I)^2|\phi_I|^2\right].
\label{classical}
\end{equation}
This is a classical potential.
Due to a quantum effect, the Chern-Simons coupling $k$ get a correction.
Because $k$ is quantized in integer units,
it cannot vary continuously.
By calculating a one loop diagram of fermion fields $\psi_I$, we find
the coupling $k(a)$ jumps by one at $a=M_I$ for each $I$\cite{3dim}.
Therefore, $\xi-\mu a$ in the classical potential (\ref{classical})
should be replaced by a function $h(a)$ satisfying
\begin{equation}
\frac{d}{da}h(a)=-\frac{g^2}{2\pi}k(a).
\label{functionh}
\end{equation}
Finally, we obtain
a quantum potential for this theory.
\begin{equation}
V(a,\phi_I)=\frac{1}{g^2}\left[\frac{1}{2}(\sum_{I=1}^{N_f}|\phi_I|^2+h(a))^2
 +\sum_{I=1}^{N_f}(a-M_I)^2|\phi_I|^2\right].
\label{quantum}
\end{equation}
\begin{figure}[htb]
\centerline{\epsfbox{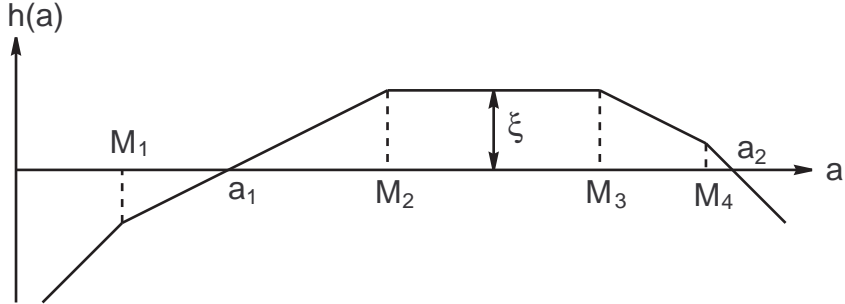}}
\caption{An example of the function $h(a)$.
We define $\xi$ as a height of a `plateau' of the function $h(a)$.
In this example, there are one Higgs vacuum
$(a=M_1,|\phi_1|=\sqrt{-h(M_1)})$ and
two Coulomb vacua $(a=a_1,\phi_I=0)$ and $(a=a_2,\phi_I=0)$.}
\label{functionh.eps}
\end{figure}
In what follows, we shall use a parameter $\xi$ as what represents
a height of a `plateau' of the function $h(a)$. (Fig.\ref{functionh.eps})
(It is not necessary for the function $h(a)$ to have a plateau.
It can be monotonically increasing or monotonically decreasing.
We assume its existence just for convenience in explanations.)

Next, let us consider vacua of this theory.
There are two kinds of vacua: `Higgs vacua' and 'Coulomb vacua'.

\paragraph{Higgs vacua}
An expansion of the potential (\ref{quantum}) in $\phi_I$ is
\begin{equation}
V(a,\phi_I)=\frac{h^2(a)}{2g^2}
           +\frac{1}{g^2}\sum_{I=1}^{N_f}(h(a)+(a-M_I)^2)|\phi_I|^2
           +{\cal O}(\phi_I^4).
\end{equation}
Roughly speaking, $\phi_I$ is tachyonic around $a=M_I$
if $h(M_I)$ is negative.
At $a=M_I$, a potential for $\phi_I$ is $V=(1/2g^2)(|\phi_I|^2+h(a))^2$
and there is supersymmetric vacuum $|\phi_I|=\sqrt{-h(M_I)}$
if $h(M_I)<0$.
This vacuum breaks the gauge symmetry.
As is mentioned in \cite{d0d6} for D0-D6 systems,
Higgs vacuum with $\phi_I\neq0$ is regarded as a true bound state
of the D2-brane and the $I$-th D8-brane.
Although it would be interesting problem how
to realize this bound state as a classical supergravity solution,
we will not argue this in this paper.

\paragraph{Coulomb vacua}
In the Coulomb branch $\phi_I=0$,
the potential is
\begin{equation}
V(a)=\frac{h^2(a)}{2g^2}.
\label{potential}
\end{equation}
There are three phases.
If $\xi<0$, the potential $V(a)$ is everywhere positive.
In this case there are supersymmetric Higgs vacua
as we mentioned above and a state on the Coulomb branch
decays into one of the Higgs vacua.
If $\xi=0$, we have a continuous set of vacua
on the plateau of the function $h(a)$.
All these vacua are supersymmetric and have an unbroken gauge symmetry.
For $\xi>0$, we have supersymmetric Coulomb vacua at $a$
satisfying $h(a)=0$.
On these vacua, both the supersymmetry and the gauge symmetry are unbroken.
The fact that only specific values of scalar field $a$ are chosen
suggests an interesting phenomenon in string theory.
It implies that the D2-brane is stabilized in the bulk at a certain distance from
the D8-branes.
In the next section, we show that the potential
$V(a)$ is reproduced as one for a D2-brane and
the stabilization
actually takes place.

\section{A classical D8-brane solution with $B$-field}
In this section, we construct a classical solution for D8-branes in $B\neq0$
and reproduce the potential (\ref{potential})
as a potential for a D2-brane in the D8-brane background.

\begin{figure}[htb]
\centerline{\epsfbox{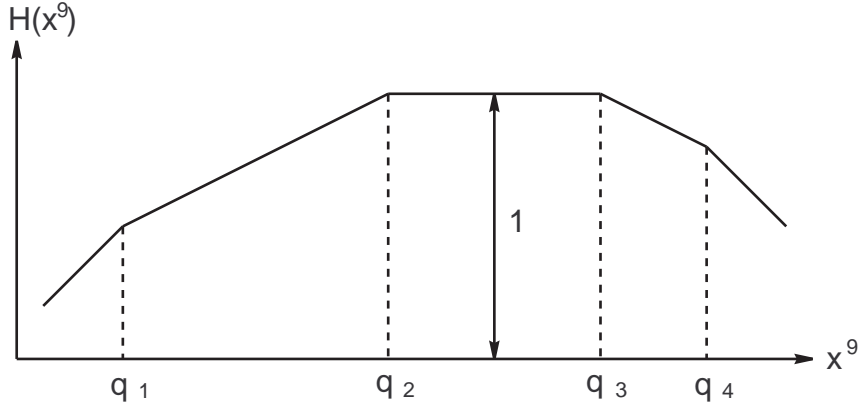}}
\caption{The harmonic function $H(x^9)$.
Each D8-brane is located at $x^9=q_I$.}
\label{harmonic.eps}
\end{figure}

We can make a D8-brane solution with a non-zero $B$-field
by T-dualizing a smeared D5-brane solution.
Let us begin with the following D5-brane solution.
\begin{eqnarray}
ds^2&=&H^{-1/2}\eta_{\mu\nu}dx^\mu dx^\nu+H^{1/2}\delta_{ij}dx^idx^j,\\
e^{\Phi}&=&g_{\rm str}H^{-1/2},\label{dilaton}\\
C_{012345}&=&\frac{2\pi}{g_{\rm str}}(2\pi l_s)^{-6}H^{-1},
\end{eqnarray}
where $\mu,\nu=0,1,\ldots,5$ and $i,j=6,7,8,9$.
The harmonic function $H$ satisfies the following Laplace equation
in the transverse directions.
\begin{equation}
\Delta_4H=-(2\pi l_s)^2g_{\rm str}\rho^{(4)},
\label{laplace}
\end{equation}
where $\rho^{(4)}$ is a D5-brane density which is now taken to be
\begin{equation}
\rho^{(4)}
=\frac{1}{(2\pi l_s)^3(-\prod_{a=1}^3\cos\theta_a)}\sum_I\delta(x^9-q_I).
\end{equation}
The $1/\prod\cos\theta_a$ factor is necessary to compensate a change of
the brane density due to the rotation which we will do next.
(We assume that one of $\theta_a$ is larger than $\pi/2$
and other two are smaller than $\pi/2$.
Therefore this density is positive.)
By integrating (\ref{laplace}) once,
we obtain
\begin{equation}
\frac{d}{dx^9}H(x^9)
=-\frac{g_{\rm str}}{2\pi l_s(-\prod_{a=1}^3\cos\theta_a)}\Lambda(x^9),
\label{harmonic}
\end{equation}
where $\Lambda(x^9)$ is a function
representing the quantized cosmological constant after the T-duality
transformation.
Because $\Lambda(x^9)$ jumps by one as $x^9$
crosses the position of each D8-brane
and is identified with the Chern-Simons coupling $k(a)$,
the function $H(x^9)$ has a similar form
to $h(a)$. (Fig.\ref{harmonic.eps})
These two functions, however, are not proportional to each other.
$H(x^9)$ must not be negative while $h(a)$ may.
Let us define $g_{\rm str}$ as an expectation value of $e^\Phi$ on the plateau
of the function $H(x^9)$.
Then, the constant part of $H(x^9)$ is fixed by (\ref{dilaton}),
such that $H(x^9)=1$ on the plateau.

Rotating this solution on $3$-$6$, $4$-$7$, and $5$-$8$ planes
by angles $\theta_1$, $\theta_2$ and $\theta_3$, respectively,
we obtain the following metric and the R-R 6-form potential.
\begin{eqnarray}
ds^2&=&H^{-1/2}(-(dx^0)^2+(dx^1)^2+(dx^2)^2)+H^{1/2}(dx^9)^2\nonumber\\
&+&\sum_{a=1}^3
   (H^{-1/2}\cos^2\theta_a+H^{1/2}\sin^2\theta_a)(dx^{a+2})^2\nonumber\\
&+&\sum_{a=1}^3
   (H^{-1/2}\sin^2\theta_a+H^{1/2}\cos^2\theta_a)(dx^{a+5})^2\nonumber\\
&+&2\sum_{a=1}^3
   \cos\theta_a\sin\theta_a(H^{-1/2}-H^{1/2})dx^{a+2}dx^{a+5},\\
C_{012678}&=&\frac{2\pi}{g_{\rm str}}(2\pi l_s)^{-6}H^{-1}\sin\theta_1\sin\theta_2\sin\theta_3.
\label{c6prime}\\
C_{012345}&=&\frac{2\pi}{g_{\rm str}}(2\pi l_s)^{-6}H^{-1}\cos\theta_1\cos\theta_2\cos\theta_3.
\label{c012345prime}
\end{eqnarray}
Other components of the R-R $6$-form potential are irrelevant to our arguments.

Compactifying the $x^6$, $x^7$ and $x^8$ directions
on a rectangular ${\bf T}^3$ with all period $2\pi l_s$
and carrying out T-duality transformation by relations
\begin{equation}
\wt G_{a+2,a+2}=G_{a+2,a+2}-\frac{G_{a+2,a+5}^2}{G_{a+5,a+5}},\quad
\wt G_{a+5,a+5}=\frac{1}{G_{a+5,a+5}},\quad
(a=1,2,3),
\label{tdual}
\end{equation}
we obtain the following metric for a D8-brane solution with
a nonzero $B$-field.
\begin{equation}
\wt G_{\mu\nu}=H^{-1/2}\eta_{\mu\nu}\quad(\mu,\nu=0,1,2),\quad
\wt G_{a+2,a+2}=\wt G_{a+5,a+5}=\frac{H^{1/2}}{F_a},\quad
\wt G_{99}=H^{1/2},
\end{equation}
where the function $F_a$ is defined by
\begin{equation}
F_a=\sin^2\theta_a+H\cos^2\theta_a.
\end{equation}
The dual dilaton field is
\begin{equation}
e^{\wt\Phi}
=e^\Phi G_{66}^{-1/2}G_{77}^{-1/2}G_{88}^{-1/2}
=g_{\rm str}H^{1/4}F_1^{-1/2}F_2^{-1/2}F_3^{-1/2}.
\end{equation}
From (\ref{c6prime}), we obtain a component of
the R-R $3$-form potential coupling to the D2-brane as
\begin{equation}
\wt C_{012}=T_{D2}H^{-1}\sin\theta_1\sin\theta_2\sin\theta_3.
\end{equation}
where $T_{D2}=1/\{(2\pi)^2l_s^3g_{\rm str}\}$
is the D2-brane tension on the plateau $H(x^9)=1$.
The zero-form R-R field strength $\wt G$ is T-dual to $G_{678}$,
which is the field strength of R-R two-form field dual to (\ref{c012345prime}).
\begin{equation}
\wt G=(2\pi l_s)^3G_{678}
     =\frac{2\pi}{g_{\rm str}}(2\pi l_s)H'
        \cos\theta_1\cos\theta_2\cos\theta_3
       =2\pi\Lambda.
\end{equation}
$\Lambda$ is certainly identified with the
cosmological constant $\wt G/2\pi$ quantized in integer units
as we mentioned above.
Using the solution we have obtained,
a potential for a D2-brane in this background is
\begin{eqnarray}
V&=&\frac{1}{(2\pi)^2l_s^3e^{\wt\Phi}}
      \sqrt{-\wt G_{00}\wt G_{11}\wt G_{22}}-\wt C_{012}\nonumber\\
 &=&T_{D2}\frac{1}{H}\left[
        \prod_{a=1}^3(\sin^2\theta_a+H\cos^2\theta_a)^{1/2}
           -\prod_{a=1}^3\sin\theta_a\right]
\label{d2pot}
\end{eqnarray}

The following relations hold between parameters in the gauge theory and ones in
string theory.
\begin{equation}
k=\Lambda,\quad
a=Tx^9,\quad
\frac{1}{g^2}=\frac{T_{D2}}{T^2}.
\label{params}
\end{equation}
Comparing (\ref{functionh}) and (\ref{harmonic}),
we obtain the following relation between $h(a)$ and $H(x^9)$.
\begin{equation}
h(a)-\xi=\left(-\prod_{a=1}^3\cos\theta_a\right)T(H(a/T)-1).
\label{hH}
\end{equation}
In order to take the field theory limit,
let us expand the potential (\ref{d2pot}) around $H=1$.
\begin{eqnarray}
\frac{V}{T_{D2}}
=A(\theta_a)+B(\theta_a)\epsilon+C(\theta_a)\epsilon^2+{\cal O}\epsilon^3.
\end{eqnarray}
where $\epsilon=1-H$.
Near the BPS point $r=0$,
the coefficients $B(\theta_a)$ and $C(\theta_a)$
are expanded in $r$ as
\begin{equation}
B(\theta_a)=r\cos\theta_1\cos\theta_2\cos\theta_3+{\cal O}(r^2),\quad
C(\theta_a)=\frac{1}{2}\cos^2\theta_1\cos^2\theta_2\cos^2\theta_3+{\cal O}(r).
\label{BC}
\end{equation}
Therefore, the potential is rewritten as
\begin{equation}
V=\frac{T_{D2}}{2T^2}(Tr+T\epsilon\cos\theta_1\cos\theta_2\cos\theta_3)^2,
\end{equation}
up to a constant.
Using (\ref{params}) and (\ref{hH}),
we can rewrite this potential in terms of variables in the field theory
except the parameter $r$.
\begin{equation}
V=\frac{1}{2g^2}(Tr+\xi-h(a))^2,
\end{equation}
This coincides with the potential (\ref{potential}) if $r$ relates to the
FI-parameter $\xi$ by
\begin{equation}
\xi=-Tr.
\end{equation}

Now we have shown that the potential in the Coulomb branch of
Maxwell-Chern-Simons-Higgs system is reproduced as
a potential for a D2-brane on the D8-brane background.
When $\xi>0$, which corresponds to $r<0$,
we have Coulomb vacua at points with $h(a)=0$.
Each of these vacua is realized by a D2-brane stabilized in the bulk
at a minimum of the potential.
Note that such a stabilization is impossible in case with
a vanishing $B$-field.
This can be seen by putting $\theta_a$ to be zero in (\ref{d2pot}).
In this case, $V\propto H^{1/2}$ and there is no stable point.

\section{Discussions}
In this paper, we showed that Maxwell-Chern-Simons-Higgs systems
are realized as gauge theories on a D2-brane in a D8-brane background
with a nonzero $B$-field.
The potential of Coulomb branch is reproduced as a potential
for a D2-brane.

There are some open questions.
We argued only the Coulomb branch in a supergravity framework.
How can we treat the Higgs branch?
One way to do it is to use a noncommutative gauge theory on D8-branes.
In the decoupling limit, $r=\sum_a(\theta_a-\pi/2)$
is proportional to $T^{-1}$.
Let us take a limit in which each $\theta_a-\pi/2$ is
proportional to $T^{-1}$.
From (\ref{bistan}), we obtain $B_{36},B_{47},B_{58}\propto T^2$
and this is what we take in order to
realize a noncommutative gauge theory on the D8-branes\cite{noncom}.
Higgs vacua are expected to be realized as noncommutative solitons
on the D8-branes.
By noncommutative parameters $\theta_{ij}$, the FI parameter $\xi$ is
represented as
\begin{equation}
-\xi
=\frac{T^2}{B_{36}}+\frac{T^2}{B_{47}}+\frac{T^2}{B_{58}}
=\frac{1}{\theta_{36}}+\frac{1}{\theta_{47}}+\frac{1}{\theta_{58}}.
\end{equation}
Although Higgs vacua are always supersymmetric,
noncommutative
soliton solutions are BPS only when $\xi=0$\cite{MPT,d0d6}.
These do not contradict to each other because
in the limit $\theta_a-\pi/2\propto T^{-1}$ the expansions
(\ref{BC}) are valid only when $\xi=r=0$.
For non-zero $\xi$,
it is not clear whether we can use solitonic solutions on the D8-branes
for the purpose of analysis of
gauge theories on the D2-branes.

A problem on supersymmetry exists for the Coulomb branch, too.
For $\xi>0$, there are supersymmetric Coulomb vacua
which are realized by stabilized D2-branes.
The D2-D8 systems,
however,
do not have any explicit unbroken supersymmetry.
\section*{Acknowledgment}
The author would like to thank K. Ohta.
Discussions with him at Summer Institute 2000 at Yamanashi, Japan was
very helpful for this work.

\end{document}